\documentclass[amsmath,amssymb,pra,twocolumn]{revtex4-2}

\usepackage{graphicx}
\usepackage{dcolumn}
\usepackage{bm}
\usepackage{hyperref}
\usepackage[mathlines]{lineno}

\usepackage{color}
\newcommand{\fracc}[2]{\frac{\textstyle{#1}}{\textstyle{#2}}}

\newcommand{\p}{\partial}

\begin{document}
\title{Spherical symmetry and nonmagnetic dielectrics in analogue models of gravity}

\author{Eduardo Bittencourt}
\email{bittencourt@unifei.edu.br}
\author{Renato Klippert}
\altaffiliation{Retired}
\affiliation{Federal University of Itajub\'a, Itajub\'a, Minas Gerais 37500-903, Brazil}

\author{Diego Renan da Silva}
\affiliation{S\~ao Paulo State University, Guaratinguet\'a, S\~ao Paulo 12516-410, Brazil}

\author{\'Erico Goulart}
\affiliation{Federal University of S\~ao Jo\~ao d'El Rei, C.A.P., Ouro Branco, Minas Gerais 36420-000, Brazil}

\date{\today}

\begin{abstract}
Nonmagnetic and spherically symmetric dielectric material media are investigated in their generality as analogue models, in the domains of geometrical optics, for a few relevant static and spherically symmetric solutions of gravitation, including black holes and wormholes. Typic and desirable gravitational features are systematically linked to the nonlinear properties of the phenomenological dielectric permittivity of the medium. From this, the limitations of such analogies are explored in terms of their corresponding geometric flexibility, dielectric configurations and causal structure.

\end{abstract}

\maketitle

\section{Introduction}
Analogue models of gravity are the easiest way for testing experimentally the kinematical aspects of any geometric model of gravity in laboratory. The first example of these models was developed by Gordon \cite{gordon} in the optical context, with further developments by De Felice \cite{deFelice:1971bgc} and Plebanski \cite{,PhysRev.118.1396}. However, the turning point came with Unruh's model on black hole evaporation in the context of fluid mechanics \cite{unruh}, where the quantum analysis became possible. Since then, several new models have been explored in the realm of condensed matter systems \cite{barcelo2005, novello2002artificial}. More recently, it has been emphasized that the optical properties of material media can be fairly easily designed nearly at will by the present day technology on materials \cite{Pendry2006,PhysRevLett.84.4184,RevModPhys.77.633}. This suggests that the optical analogue models could be favoured as natural candidates for mimicking the behavior of gravitational systems from the experimental of view.

The present work aims to draw the basic lines to guide material designers, from the perspective that each kind of matter can be used to produce the desired effective gravitational potential. In order to obtain definite results of broad significance, some simplification is needed; the discussion applies to nonmagnetic isotropic dielectric media, which are used to model some properties of static and spherically symmetric solutions of the theory of gravitation.

For completeness, Sec.\ \ref{2} reviews the theory of optical analogue models through the generalized Fresnel eigenvalue problem up to obtaining the ordinary and extraordinary modes of light propagation and the two corresponding optical analogue metrics. In Sec.\ \ref{Analogue} the general analogue models are obtained and the corresponding electromagnetic quantities are provided and compared with two paradigmatic gravitational models: the black hole and the wormhole models. Section \ref{hypandstruc} discusses the causality issues related to Cauchy developments. The effective light cones were constructed in order to provide one possible physical interpretation for the optical analogue metrics.

Notation is such that Latin indices $i,j,l,\ldots$ run from $1$ to $3$ and repeated indices are summed upon. Partial derivatives are denoted as $\p_i=\p/\p x^i$, $\p_t=\p/\p t$, and $\eta_{ijl}$ is the completely skew-symmetric Levi-Civita symbol with $\eta_{123}=1$.

\section{Geometric optics inside a nonlinear dielectric}
\label{2}
To begin with, recall that Maxwell equations inside a material medium can be written, in a Cartesian coordinate system, as
\begin{equation}
\label{max_diel}
\begin{array}{l}
\p_iD_i=\rho,\qquad \eta_{ijl}\p_jE_l=-\p_t B_i,\\[1ex]
\p_iB_i=0,\qquad \eta_{ijl}\p_jH_l= J_i + \p_t D_i,
\end{array}
\end{equation}
where $D_i$ and $H_i$ describe the field excitations and $E_i$ and $B_i$ stand for the corresponding electromagnetic field strengths. The nonlinear behavior of the medium is taken into account by replacing the usual linear constitutive relations by the following nonlinear ones:
\begin{eqnarray}
D_i=\epsilon_0E_i + P_i,&\quad\longrightarrow\quad& D_i = \epsilon_{ij}  E_j,\label{cons-rel1}\\[2ex]
H_i=\frac{1}{\mu_0}B_i-M_i,&\quad\longrightarrow\quad& H_i= \mu^{{}_{-1}}_{ij} B_j.\label{cons-rel2}
\end{eqnarray}
Here, $P_i$ is the polarization, $M_i$ is the magnetization, and $\epsilon_0$ and $\mu_0$ are the vacuum dielectric constants. In general cases, the coefficients of permittivity  $\epsilon_{ij}$ and the inverse permeability $\mu^{{}_{-1}}_{ij}$ both can depend on the fields $E_i$ and $B_i$. Note that extra terms accounting for the electric/magnetic excitation due to magnetic/electric field strength can both be included, but we shall not deal with this here (see Ref.\ \cite{perlick2011} for details).

In the regime of geometric optics, the field strengths, the charge $\rho$ and the current $J_i$ are continuous, but their derivatives may have a finite step through the wavefront $\Sigma_t(x_i)= \mbox{const.}$, for a given instant of time \(t\). According to Hadamard's theorem~\cite{hadamard,zakharov}, \(\left[E_i\right]_\Sigma=0=\left[B_i\right]_\Sigma\) implies
\begin{equation}
\label{hada}
\begin{array}{l}
\left[\p_t E_i\right]_{\Sigma_t}=-\omega\, e_i, \quad \left[\p_j E_i\right]_{\Sigma_t}=e_i\,k_j,\\[1ex]
\left[\p_t B_i\right]_{\Sigma_t}=-\omega\, b_i, \quad \left[\p_j B_i\right]_{\Sigma_t}=b_i\,k_j,\\[1ex]
\end{array}
\end{equation}
where the symbol $\left[ f \right]_{\Sigma_t}(p) \equiv \lim_{\delta\to0^+} [f(p_+) - f(p_-)]$
indicates how the step of the electromagnetic quantities are evaluated on $\Sigma_t$, with $p\in\Sigma_t$ and $p_{\pm}\in\Sigma_{t\pm\delta}$ such that \(\lim_{\delta\to0^+} p_{\pm}=p\). We also denote the wave frequency by $\omega$ and the wave vector by $k_i$. The polarization modes of the electric and magnetic components of the light rays are represented by $e_i$ and $b_i$, respectively.

Once applied to the Maxwell equations (\ref{max_diel}), the procedure described above yields a linearly polarized wave with the magnetic polarization being given by $\omega b_i =\eta_{ijl}k_j e_l$ , while the electric polarization $e_i$ satisfies the eigenvalue problem \cite{DELORENCI200661}
\begin{equation}
\label{eig_val}
Z_{ij} e_j=0,
\end{equation}
where $Z_{ij}$ represent the components of the Fresnel matrix \cite{goulart2008}. The existence of nontrivial solutions for this eigenvalue problem is equivalent to the requirement
\begin{equation}
\det[Z_{ij}]=0.
\label{fresnel_cond}
\end{equation}
From this equation, the dispersion relation associated with the light rays can then be obtained and the nonlinear behavior of the system is encoded in the dielectric parameters.

It is worth to mention that the vectors $E_i$ and $B_i$ represent the total field, that is, the composition of an external field plus a wave field. As long the limit of geometric optics is valid, the fields associated with the propagating waves are much weaker than the controllable fields---those produced by a given source distribution in the medium or by imposed external fields. Thus, wave fields were consistently neglected in these sums.

The majority of the optical media respond linearly to external magnetic fields. This means that the permeability can be considered as $\mu^{{}_{-1}}_{ij}=(1/\mu_0)\delta_{ij}$, where $\mu_0$ is the vacuum permeability and $\delta_{ij}$ is the Kronecker delta, which is either $1$ for $i=j$ or $0$ otherwise. Besides, we shall consider the simplest case of an isotropic nonlinear electric medium such that the permittivity matrix is the same for all spatial directions and depends only on the norm $E=\sqrt{E_i E_i}$ of the electric field, that is, $\epsilon_{ij}=\epsilon(E) \delta_{ij}$. In this case, the Fresnel matrix is given by \cite{PhysRevA.95.033826}
\begin{equation}
\label{red_fresnel}
Z_{ij} = \epsilon \delta_{ij} + \frac{\epsilon'}{E} E_{i}E_{j}-\frac{1}{\mu_0 v^2}(\delta_{ij}-\hat{k}_i\hat{k}_j),
\end{equation}
where $\epsilon'=d\epsilon/dE$, $v=\omega/k$ is the phase velocity and $\hat{k}_i=k_i/k$, with $k=\sqrt{k_i k_i}$.

Straightforward calculations show that Eq.\ (\ref{fresnel_cond}) can be written as a polynomial equation for the phase velocity and a simple manipulation of such equation casts it as a pair of equations of the form \(g^{\mu\nu}k_\mu k_\nu=0\), from which we can read out the effective optical metrics \cite{edu2012,Bittencourt_2016}. With Greek indices running from $0$ to $3$ and the Minkowski background metric in Cartesian coordinates taking the form  $[\eta_{\mu\nu}] = {\rm diag}(-1,+1,+1,+1)$, these effective metric matrices can be written as
\begin{eqnarray}
&g^{(+)}_{\alpha\beta}=\gamma_{\alpha\beta}+\left(1-\fracc{1}{\mu_0\epsilon}\right)v_\alpha v_\beta
\label{eff_met+},\\[2ex]
&\hspace{-0.3cm}g^{(-)}_{\alpha\beta}=\gamma_{\alpha\beta}+\left[1-\fracc{1}{\mu_0(\epsilon+\epsilon'E)}\right]v_\alpha v_\beta-\fracc{\epsilon'E}{\epsilon+\epsilon'E}\hat l{}_\alpha\hat l{}_\beta
\label{eff_met-}\label{opt_up},
\end{eqnarray}
where $\gamma_{\alpha\beta}$ denotes the Minkowski metric in an arbitrary coordinate system,  $v^{\mu}=\delta^{\mu}_0$ represents the class of observers comoving with the laboratory, that is, it measures the electric and magnetic components of the electromagnetic fields, and $\hat l{}^\alpha$ is a unit vector pointing along the direction of the electric field. The effective metric $g^{(+)}_{\alpha\beta}$ is the well-known Gordon metric \cite{gordon,edu2012} and it easily is recognizable by its isotropicity. This metric is responsible for describing the propagation of the ordinary modes. On the other hand, the optical metric $g^{(-)}_{\alpha\beta}$ does depend on the nonlinear behavior of the dielectric and it shall be responsible for the propagation of the extra-ordinary modes. This is the metric we are interested in.

It is easy to see that, in the particular case of vacuum, where the dielectric coefficients are $\epsilon_0$ and $\mu_0$, these two metrics degenerate to a unique optical metric $g_{\mu\nu}$ given by the Minkowski one. The relevance of these optical metrics lies on the fact that the integral curves of the wave vector $k_\mu$ correspond to null geodesic curves in the artificial spacetime endowed with a geometry given by the effective optical metric \cite{delorenci2002,delorenci2002b}.
As far as solely kinematic aspects of the gravitation are considered, the effective geometry could be compared with the metric of the curved spacetime. Thus, the sort of phenomena predicted in gravitational systems could be scrutinized in the realm of optics inside material media. There are many applications of this analogy including tests of bending of light, cosmological models in terrestrial laboratories, or the probe of predictions in quantum gravity phenomenology \cite{barcelo2005}.

\section{Static and spherically symmetric optical metrics}
\label{Analogue}
Let us consider a dielectric material with the features described above, subjected to an external radial electric field and without external magnetic fields. For convenience, we choose coordinates adapted to the observers such that the four-velocity is $v^{\alpha}=\delta^{\alpha}_0$. Then, the static and spherically symmetric situation we are dealing with admits solely a non-vanishing charge density $\rho$. Therefore, Maxwell equations (\ref{max_diel}) essentially reduce to
\begin{equation}
\frac{\partial_r(r^2\epsilon E)}{r^2}=\rho.
\label{max}
\end{equation}
Thus, the effective geometry Eq.~(\ref{opt_up}) expressed in spher\-ical-like coordinates $(t,r,\theta,\phi)$ reads
\begin{equation}
g_{\alpha\beta}={\rm diag}\left(-\fracc{1}{\mu_0(\epsilon+\epsilon'E)},\,\frac{\epsilon}{\epsilon+\epsilon'E},\,r^2,\,r^2\sin^2\theta\right)
\label{diag}.
\end{equation}
This expression allows one to seek for analogue static and spherically symmetric metrics of the following form
\begin{equation}
g_{\alpha\beta}={\rm diag}\left(-e^{2\Phi(r)},\,\fracc{1}{1-\frac{b(r)}{r}},\,r^2,\,r^2\sin^2\theta\right).
\label{st-sp}
\end{equation}
Inspired by the Morris-Thorne \textit{Ansatz} concerning wormholes \cite{morris}, the redshift function $\Phi(r)$ and the shape function $b(r)$ are enough to describe any static and spherically symmetric spacetime in a unified form. In particular, the well-known solutions of Einstein equations such as Schwarzschild, Reissner-Nordstr\"om, de-Sitter as well as the class of Morris-Thorne wormholes. The matching of Eqs.\ (\ref{diag}) and (\ref{st-sp}) gives the two independent equations
\begin{eqnarray}
\mu_0(\epsilon+\epsilon'E)=e^{-2\Phi}, \label{eq_epsilon1}\\
\frac{\epsilon+\epsilon'E}{\epsilon}=1-\frac{b}{r}.\label{eq_epsilon2}
\end{eqnarray}
Once combined, these conditions provide the radial dependence of the dielectric permittivity, that is
\begin{equation}
\label{epsilon_r}
\epsilon(r)=\epsilon_0\,e^{-2\Phi}\left(1-\frac{b}{r}\right)^{-1},
\end{equation}
where we used the identification $\mu_0^{-1}=\epsilon_0$, since the speed of light in vacuum is set to unity. Substitution of Eq.\ (\ref{epsilon_r}) back in Eq.\ (\ref{eq_epsilon1}) yields the electric field in terms of the radial coordinate, as follows
\begin{equation}
E(r)=E_0\left|\frac{r}{b}-1\right|\exp\left(2\int\frac{r}{b}\,\frac{d\Phi}{dr}dr\right),
\label{E_r}
\end{equation}
where $E_0$ is an integration constant. This equation is valid for any signs of $r-b$ and $b$. However, we are mainly interested in the case $r-b\geq0$, in order to preserve the Lorentzian signature of metric in Eq. (\ref{st-sp}).

Finally, the expressions of the electric displacement $D=\epsilon E$ and the charge density $\rho$ in terms of $r$ read as
\begin{eqnarray}
D&=&\epsilon_0 E_0\,\fracc{r}{b}\exp{\left[2\int\left(\fracc{r}{b}-1\right)\,\fracc{d\Phi}{dr}dr\right]}
\label{D-r},\\
\rho&=&\epsilon_0 E_0\,\left[\fracc{3}{b} - \fracc{r}{b^2}\fracc{db}{dr} + 2\,\fracc{d\Phi}{dr}\left(\fracc{r}{b}-1\right) \right]\times\nonumber\\
&&\times\exp{\left[2\int\left(\fracc{r}{b}-1\right)\,\fracc{d\Phi}{dr}dr\right]}.
\label{rho}
\end{eqnarray}
Therefore, for any choice of the functions $\Phi(r)$ and $b(r)$ describing the gravitational metric to be mimicked, the corresponding analogue model can be constructed, at least theoretically, via equations (\ref{epsilon_r})-(\ref{rho}). In what follows, two important cases of static and spherically symmetric geometries in the context of general relativity are considered. Some possible experimental limitations for reproducing such metrics in laboratory are also discussed latter, by assuming a nonmagnetic and nonlinear isotropic material.

\subsection{Black hole analogue models}
First, consider the case in which the functions \(\Phi(r)\) and \(b(r)\) correlate as
\begin{equation}
\label{phi_bh}
\Phi(r)=\frac{1}{2}\ln\left(1-\frac{b(r)}{r}\right).
\end{equation}
This choice leads to a gravitational metric of the form given in Eq.\ (\ref{st-sp}) where the condition $g_{00}=-1/g_{11}$ holds. This condition includes all black hole metrics like Schwarzschild, Kottler and Reissner-Nordstr\"om (De Sitter and anti-De Sitter spacetimes are also solutions of this, but without event horizons).

Accordingly, when Eqs.\ (\ref{epsilon_r})-(\ref{rho}) are restricted to the case of Eq.\ (\ref{phi_bh}), the corresponding analogue model presents the following electromagnetic configuration
\begin{equation}
\label{eq_em_bh}
\begin{array}{lcl}
\epsilon(r)&=&\epsilon_0\left(1-\fracc{b}{r}\right)^{-2},  \\[1ex]
E(r)&=&E_0\left(\fracc{r}{b}-1\right)^{2},  \\[1ex]
D(r)&=& \fracc{\epsilon_0 E_0 r^2}{b^2}, \\[1ex]
\rho(r)&=& \fracc{2\epsilon_0 E_0 r^2}{b^2}\left(\fracc{2}{r}-\fracc{1}{b}\fracc{db}{dr}\right) .
\end{array}
\end{equation}
In particular, for this case one has
\begin{equation}\label{perm1}
\epsilon(E)=\epsilon_0\left(\sqrt{\frac{E_0}{E}}\pm1\right)^2.
\end{equation}
Thus, the permittivity diverges as $E$ approaches zero. Therefore, we have a large permittivity for small values of $E$, maintaining always the electric displacement finite (see Eqs.\ \ref{eq_em_bh}). This cannot be achieved by using materials one can find in nature, but possibly with short-band human made media like metamaterials. Besides, the physically reasonable expression for $\epsilon(E)$ is the one with the plus sign, avoiding the situation $\epsilon|_{E_0}=0$.

\subsection{Wormhole analogue models}
The (intuitive) conditions for the existence of a wormhole without an event horizon are that $\Phi(r)$ is finite everywhere and $b(r_0)=r_0$ for some $r=r_0>0$ \cite{morris}. This implies that the permittivity from Eq.\ (\ref{epsilon_r}) diverges at the throat, while the electric field (\ref{E_r}) goes to zero, at least as far as the argument of the exponential remains finite. For the sake of illustration, take the special class of Morris-Thorne wormholes with zero redshift function. Thus,
\begin{equation}
\begin{array}{lcl}
\epsilon(r)&=&\epsilon_0\left(1-\fracc{b}{r}\right)^{-1},  \\[1ex]
E(r)&=&E_0\left(\fracc{r}{b}-1\right),  \\[1ex]
D(r)&=& \fracc{\epsilon_0 E_0 r}{b}, \\[1ex]
\rho(r)&=& \epsilon_0 E_0\,\left(\fracc{3}{b} - \fracc{r}{b^2}\fracc{db}{dr}\right) .
\end{array}
\end{equation}
The elimination of the variable \(r\) from the first two relations above yields
\begin{equation}
\epsilon(E)=\epsilon_0\left(\frac{E_0}{E}+1\right).
\end{equation}
Although the equations are slightly different from the previous case, the conclusion is quite similar: a divergent permittivity and a finite electric displacement as the electric field goes to zero, which happens at the throat at least. Again, metamaterials would be required in order to reproduce this sort of geometries.

\subsection{Power-law media}
The previous subsections have shown that \textit{any} gravitational black-hole or wormhole (with zero redshift function) cannot be reproduced in laboratory by conventional materials - those media whose electric permittivity is proportional to a positive power of the magnitude of the electric field, i.e. $\epsilon\propto E^{\alpha}$ with $\alpha>0$. Then, the natural question raises: what kind of metrics can be produced with conventional media? Thus, consider a power law relation between the electric permittivity and the electric field. With this assumption we cover the well-known Pockels and Kerr effects, first and second order corrections in $E$, respectively, to the electric susceptibility of a given medium.

Thus, from the substitution of
\begin{equation}
\epsilon(E)=\epsilon_0\left(\fracc{E}{E_0}\right)^{\alpha},
\end{equation}
with $\alpha>0$ in Eqs.\ (\ref{epsilon_r}) and (\ref{E_r}), it follows that
\begin{equation}
\label{phi_power_law}
\Phi=-\frac{1}{2}\ln\left(1-\frac{b}{r}\right),
\end{equation}
for any $\alpha>0$. Please, note a sign on comparing this with Eq.\ (\ref{phi_bh}). Returning back to Eqs.\ (\ref{epsilon_r}) and (\ref{E_r}), one finds out that $\epsilon$ and $E$ are actually constant and that the effective metric reduces to the flat Minkowski spacetime. Once again, nontrivial effective metrics that are static and spherically symmetric demand a more ellaborated relationship between $\epsilon$ and $E$.

\subsection{Neutral dielectric media}
Finally, assume that the charge density vanishes identically, such that the profile of the electric displacement is directly obtained from Gauss law, that is
\begin{equation}
D=\frac{k}{r^2},
\end{equation}
where $k$ is an integration constant. The comparison of this pattern with the product $\epsilon E$ given by Eqs.\ (\ref{epsilon_r}) and (\ref{E_r}) yields a first order differential equation involving $\Phi(r)$ and $b(r)$, which can be solved, leading to
\begin{equation}
\Phi(r)=-\frac{1}{2}\ln\left(1-\frac{b}{r}\right) + \ln\frac{r}{r_0} - \int\frac{dr}{(r-b)}.
\end{equation}
The last two terms on the right-hand side of this relation distort the solution in Eq.\ (\ref{phi_power_law}), yielding a nontrivial effective metric. The dielectric parameter and the electric field strength are, respectively, given by
\begin{equation}
\begin{array}{lcl}
\epsilon(r)&=&\epsilon_0\left(\fracc{r_0}{r}\right)^2\exp\left(2\int\frac{dr}{(r-b)}\right),\\[2ex]
E(r)&=&E_0\exp\left(-2\int\frac{dr}{(r-b)}\right).
\end{array}
\end{equation}
From this, it is clear that $k=\epsilon_0 E_0 r_0^2$. Moreover, the permittivity is $\epsilon=\epsilon_0 E_0 r_0^2\, r^{-2}E^{-1}$. The actual dependence of \(\epsilon\) on \(E\) requires the explicit behavior of \(E(r)\) to be algebraically inverted, which in turns demands the specification of the shape function \(b(r)\).

\section{Hyperbolicity and causal structure}
\label{hypandstruc}

In this Section, the interconnected issues of hyperbolicity and causal structure for the models described so far are briefly discussed. Hyperbolicity lies at the roots of physics, since it amounts to the predictable power of the theory by asserting that solutions exist, are unique and depend continuously on the initial data. For physical models described by quasi-linear partial differential equations with constraints, as is the case under investigation, additional care is needed since the principal part of the equations will depend explicitly on the electromagnetic fields. Roughly speaking, this means that the Cauchy problem may be well-posed for some field intensities while ill-posed for others. Needless to say, if hyperbolicity was to be violated, one could have dispersion relations endowed with pathologies, which would compromise the propagation of waves assumed at the very beginning of our analysis. These pathologies are often associated with non-real eigenvalues of the characteristic polynomial (evanescent modes) or cones of influence with non-convex topologies.

The hyperbolicity of Maxwell equations with local, but otherwise arbitrary, constitutive laws has been analyzed in \cite{PhysRevD.83.044047,perlick2011}. In particular, one inescapable condition for well-posedness to hold for Eqs.\ (\ref{max_diel})-(\ref{cons-rel2}) is that, at each spacetime point, the homogeneous multivariate polynomial of fourth order
\begin{equation}
P(\xi)\equiv (g_{\alpha\beta}^{(+)}\xi^{\alpha}\xi^{\beta})(g_{\alpha\beta}^{(-)}\xi^{\alpha}\xi^{\beta})
\end{equation}
is hyperbolic with respect to a direction in the tangent space \cite{10.2307/24900665,Beig2006}. In other words, $\exists\ w^{\mu}\in T_{p}M$ such that $P(w)>0$ and the map $\lambda\mapsto P(u+\lambda w)$, itself a univariate polynomial of fourth order, has only real roots $\lambda_{i}$, for all $u^{\mu}\in T_{p}M$. Geometrically, hyperbolicity in the direction of $w^{\mu}$ is the requirement that every line parallel to $w^{\mu}$ in $T_{p}M$ intersects the null cones of $g_{\alpha\beta}^{(+)}$ and $g_{\alpha\beta}^{(-)}$ at exactly four points (counting multiplicities). Therefore, one obtains the conditions
\begin{enumerate}
\item{the effective metrics $g_{\alpha\beta}^{(+)}$ and $g_{\alpha\beta}^{(-)}$ both must have a Lorentzian signature equal to $(-,+,+,+)$;}
\item{ the set of future-directed time-like vectors with respect to both metrics must form a unique connected, open, convex set.}
\end{enumerate}
Needless to say, if the above requirements are satisfied, the causal structure of the theory will be well-behaved: wave excitations or discontinuities over any background configuration will propagate in a predictable way with a finite velocity of propagation.

Now, a direct calculation using Eqs.\ (\ref{eff_met-}) shows that the effective metrics will be Lorentzian, whenever the following inequalities are satisfied
\begin{equation}
g^{(+)}=\frac{\gamma}{\mu_{0}\epsilon}<0,\qquad g^{(-)}=\frac{\gamma}{\mu_{0}\epsilon(1+\frac{\epsilon'E}{\epsilon})^{2}}<0.
\end{equation}
where $g^{(\pm)}=\det\left(g_{\alpha\beta}^{(\pm)}\right)$ and $\gamma\equiv \mbox{det}(\gamma_{\alpha\beta})$. The latter imply that, in order to preserve the correct signature, the following two conditions must be fulfilled:
\begin{equation}
\epsilon>0,\quad\quad\quad \epsilon+\epsilon'E> 0.
\end{equation}
These two conditions are automatically satisfied by all spherically symmetric models described so far. That condition 2 above is also fulfilled is immediate. Since the class of observers comoving with the laboratory, $v^{\mu}$, is time-like with respect to both effective metrics then, by continuity, there will be an open set of time-like vectors containing $v^{\mu}$ which is necessarily connected, open and convex, as required. Hence, these models are not plagued by mathematical inconsistencies.

\begin{figure}[ht]
    \centering
    \includegraphics[scale=0.38]{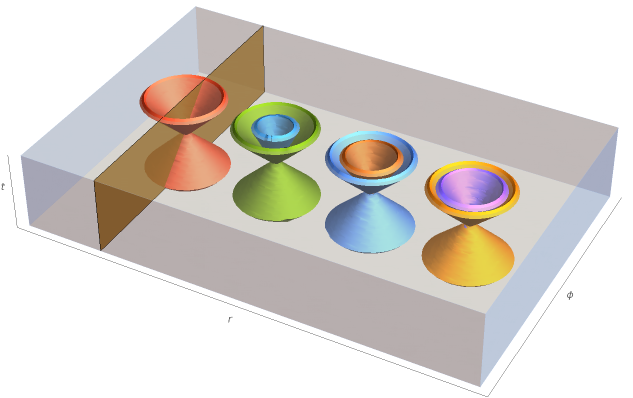}
    \caption{Null cones of $g^{(+)}_{\mu\nu}$ are inside the Minkowski light cones, shown as a function of $r$. The vertical plane on left represents the \textit{worldsheet} of the horizon, i.e. $r=r_{H}$, for which the effective light cones degenerate. As the $r$-coordinate grows, the vertices of the other light cones are placed at $2r_H, 3r_H$ and $4r_H$, respectively. Also, $\mu_0^{-1}=\epsilon_0=1=E_0$ were adopted for simplicity. Notice the isotropic pattern of the propagation.}
    \label{fig:ortho}
\end{figure}

\begin{figure}[ht]
    \centering
    \includegraphics[scale=0.38]{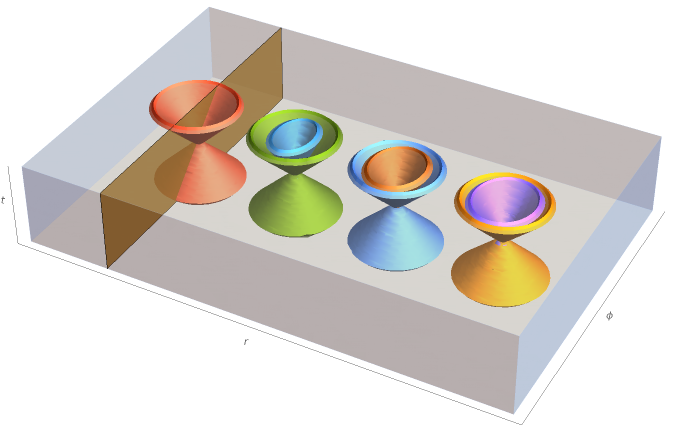}
    \caption{Null cones of the effective metric $g^{(-)}_{\mu\nu}$ also are inside the Minkowski light cones, here shown as a function of $r$. All numerical choices are the same as in Fig.\ \ref{fig:ortho}. The anisotropic pattern is due to the term $\epsilon'E$ in the effective metric. Since this term vanishes for large values of $r$, the effective cones gradually approach Minkowski light cones from inside.}
    \label{fig:ortho2}
\end{figure}

Another interesting feature of the effective metrics, Eqs.\ (\ref{eff_met+})-(\ref{eff_met-}), is that either their null cones coincide or they share two (and only two) common directions. Indeed, in Cartesian coordinates adapted to the laboratory frame i.e., $v^{\mu}=\delta^{\mu}_{0}$ and $\gamma_{\mu\nu}=\eta_{\mu\nu}$, we may write $\xi^{\mu}=(\xi^{0},\vec{\xi})$ and the corresponding null cone conditions read as

\begin{equation}\label{nullcones}
(\xi^{0})^{2}-\mu_{0}\epsilon (\vec{\xi}\cdot\vec{\xi})=0,\qquad (\xi^{0})^{2}-\mu_{0}(\epsilon+\epsilon'E \mbox{sin}^{2}\psi )(\vec{\xi}\cdot\vec{\xi})=0 \end{equation}
with $\psi$ denoting the angle between $\vec{\xi}$ and the electric field $\vec{E}$. Therefore, if $\epsilon'E=0$, the cones are precisely the same, whereas if $\epsilon'E\neq 0$, these cones agree along the directions of $\pm\vec{E}$. This means that, in a static and spherically symmetric situation, it would be impossible to distinguish between the two effective geometries, and consequently between the ordinary and the extra-ordinary light rays, by solely studying the behavior of radial light propagation.

For a concrete example, consider the causal structure in the simple case of a Schwarzschild geometry, for which $b(r)=r_{H}$. A direct computation using Eqs.\ (\ref{eq_em_bh}) gives the following electromagnetic quantities
\begin{equation}
\begin{array}{lcl}
\epsilon(r)&=&\epsilon_0\left(1-r_{H}/r\right)^{-2},  \\[1ex]
E(r)&=&E_0\left(r/r_{H}-1\right)^{2},  \\[1ex]
D(r)&=& \epsilon_0 E_0 \left(r/r_{H}\right)^{2}, \\[1ex]
\rho(r)&=& 4\epsilon_0 E_0 (r/r^{2}_{H}),
\end{array}
\end{equation}
which are finite in the regime $r_{H}<r<\infty$. Thus, one concludes that the electric permittivity is a positive definite, monotonically decreasing function, diverging as $r\rightarrow r_{H}$ and approaching $\epsilon_{0}$ as $r\rightarrow \infty$. Conversely, the electric field intensity vanishes as $r\rightarrow r_{H}$ and diverges as $r\rightarrow\infty$. Also, by noticing that
\begin{equation}
\epsilon'E=\left(\frac{d\epsilon}{dr}\right)\left(\frac{dE}{dr}\right)^{-1}E=\epsilon_0\frac{r_H}{r}\left(1-\fracc{r_H}{r}\right)^{-2}
\end{equation}
one easily shows that the \textit{anisotropic term}, given by $\epsilon'E$,
vanishes as $r\rightarrow\infty$, as expected. Therefore, the effective null cones, Eqs.\ (\ref{nullcones}), coincide at spatial infinity and increasingly disagree as the radial coordinate is gradually decreased. Nevertheless, the two cones always intersect for rays propagating along the radial direction, as discussed before. Clearly, both cones coincide with the Minkowski light cone for $r\gg r_{H}$, and they degenerate at $r=r_{H}$, which is avoided here by considering only the exterior region.

The qualitative behavior of the light cones for both effective metrics are depicted in Figs.\ (\ref{fig:ortho})-(\ref{fig:ortho2}) for different values of the radii, that is, $r/r_H=1,2,3,4$. Figure~(\ref{fig:ortho}) displays the effectively causal cones defined by $g_{\mu}^{(+)}$. The effective light cone lies inside the Minkowski background causal cone, and approaches it as $r$ goes to infinity. The isotropic feature of the effective cones is evident. In the case of \(g_{\mu}^{(-)}\) in Fig.\ (\ref{fig:ortho2}), the same qualitative behavior for the effective light cones of $g_{\mu\nu}^{(-)}$ for the same values of the parameter, despite their anisotropy.

\section{Concluding Remarks}
Most of the literature on analogue models present some interesting properties of a certain kind of material medium, without actually providing what is the specific feature of the material which allows the obtained behavior. The present work is a first step to construct a global picture, by discussing what it can be obtained, and what cannot, from these properties. In particular, it was shown that most of the non standard optical properties of non magnetic isotropic dielectrics demand a medium whose electric permittivity diverges in the limit of $E \rightarrow 0$. The discussion of the causal structure provides the basic requirements on the effective metrics in order to mimic some reliable spacetime geometry.

\begin{acknowledgments}
DRS would like to thank \textit{Coordenação de Aper\-feiçoamento de Pessoal de Nível Superior} (CAPES) for the financial support.
\end{acknowledgments}

\bibliography{ref}

\providecommand{\noopsort}[1]{}\providecommand{\singleletter}[1]{#1}%
\begin{thebibliography}{23}%
\makeatletter
\providecommand \@ifxundefined [1]{%
 \@ifx{#1\undefined}
}%
\providecommand \@ifnum [1]{%
 \ifnum #1\expandafter \@firstoftwo
 \else \expandafter \@secondoftwo
 \fi
}%
\providecommand \@ifx [1]{%
 \ifx #1\expandafter \@firstoftwo
 \else \expandafter \@secondoftwo
 \fi
}%
\providecommand \natexlab [1]{#1}%
\providecommand \enquote  [1]{``#1''}%
\providecommand \bibnamefont  [1]{#1}%
\providecommand \bibfnamefont [1]{#1}%
\providecommand \citenamefont [1]{#1}%
\providecommand \href@noop [0]{\@secondoftwo}%
\providecommand \href [0]{\begingroup \@sanitize@url \@href}%
\providecommand \@href[1]{\@@startlink{#1}\@@href}%
\providecommand \@@href[1]{\endgroup#1\@@endlink}%
\providecommand \@sanitize@url [0]{\catcode `\\12\catcode `\$12\catcode
  `\&12\catcode `\#12\catcode `\^12\catcode `\_12\catcode `\%12\relax}%
\providecommand \@@startlink[1]{}%
\providecommand \@@endlink[0]{}%
\providecommand \url  [0]{\begingroup\@sanitize@url \@url }%
\providecommand \@url [1]{\endgroup\@href {#1}{\urlprefix }}%
\providecommand \urlprefix  [0]{URL }%
\providecommand \Eprint [0]{\href }%
\providecommand \doibase [0]{https://doi.org/}%
\providecommand \selectlanguage [0]{\@gobble}%
\providecommand \bibinfo  [0]{\@secondoftwo}%
\providecommand \bibfield  [0]{\@secondoftwo}%
\providecommand \translation [1]{[#1]}%
\providecommand \BibitemOpen [0]{}%
\providecommand \bibitemStop [0]{}%
\providecommand \bibitemNoStop [0]{.\EOS\space}%
\providecommand \EOS [0]{\spacefactor3000\relax}%
\providecommand \BibitemShut  [1]{\csname bibitem#1\endcsname}%
\let\auto@bib@innerbib\@empty
\bibitem [{\citenamefont {Gordon}(1923)}]{gordon}%
  \BibitemOpen
  \bibfield  {author} {\bibinfo {author} {\bibfnamefont {W.}~\bibnamefont
  {Gordon}},\ }\bibfield  {title} {\bibinfo {title} {Zur lichtfortpflanzung
  nach der relativitätstheorie},\ }\href@noop {} {\bibfield  {journal}
  {\bibinfo  {journal} {Annalen der Physik}\ }\textbf {\bibinfo {volume}
  {72}},\ \bibinfo {pages} {421} (\bibinfo {year} {1923})}\BibitemShut
  {NoStop}%
\bibitem [{\citenamefont {de~Felice}(1971)}]{deFelice:1971bgc}%
  \BibitemOpen
  \bibfield  {author} {\bibinfo {author} {\bibfnamefont {F.}~\bibnamefont
  {de~Felice}},\ }\bibfield  {title} {\bibinfo {title} {On the gravitational
  field acting as an optical medium},\ }\href@noop {} {\bibfield  {journal}
  {\bibinfo  {journal} {Gen. Rel. and Grav.}\ }\textbf {\bibinfo {volume}
  {2}},\ \bibinfo {pages} {1396} (\bibinfo {year} {1971})}\BibitemShut
  {NoStop}%
\bibitem [{\citenamefont {Plebanski}(1960)}]{PhysRev.118.1396}%
  \BibitemOpen
  \bibfield  {author} {\bibinfo {author} {\bibfnamefont {J.}~\bibnamefont
  {Plebanski}},\ }\bibfield  {title} {\bibinfo {title} {Electromagnetic waves
  in gravitational fields},\ }\href {https://doi.org/10.1103/PhysRev.118.1396}
  {\bibfield  {journal} {\bibinfo  {journal} {Phys. Rev.}\ }\textbf {\bibinfo
  {volume} {118}},\ \bibinfo {pages} {1396} (\bibinfo {year}
  {1960})}\BibitemShut {NoStop}%
\bibitem [{\citenamefont {Unruh}(1981)}]{unruh}%
  \BibitemOpen
  \bibfield  {author} {\bibinfo {author} {\bibfnamefont {W.~G.}\ \bibnamefont
  {Unruh}},\ }\bibfield  {title} {\bibinfo {title} {Experimental black-hole
  evaporation?},\ }\href {https://doi.org/10.1103/PhysRevLett.46.1351}
  {\bibfield  {journal} {\bibinfo  {journal} {Phys. Rev. Lett.}\ }\textbf
  {\bibinfo {volume} {46}},\ \bibinfo {pages} {1351} (\bibinfo {year}
  {1981})}\BibitemShut {NoStop}%
\bibitem [{\citenamefont {Barcel\'o}\ \emph {et~al.}(2005)\citenamefont
  {Barcel\'o}, \citenamefont {Liberati},\ and\ \citenamefont
  {Visser}}]{barcelo2005}%
  \BibitemOpen
  \bibfield  {author} {\bibinfo {author} {\bibfnamefont {C.}~\bibnamefont
  {Barcel\'o}}, \bibinfo {author} {\bibfnamefont {S.}~\bibnamefont
  {Liberati}},\ and\ \bibinfo {author} {\bibfnamefont {M.}~\bibnamefont
  {Visser}},\ }\bibfield  {title} {\bibinfo {title} {Analogue gravity},\
  }\href@noop {} {\bibfield  {journal} {\bibinfo  {journal} {Living Reviews in
  Relativity}\ }\textbf {\bibinfo {volume} {8}},\ \bibinfo {pages} {12}
  (\bibinfo {year} {2005})}\BibitemShut {NoStop}%
\bibitem [{\citenamefont {Novello}\ \emph {et~al.}(2002)\citenamefont
  {Novello}, \citenamefont {Visser},\ and\ \citenamefont
  {Volovik}}]{novello2002artificial}%
  \BibitemOpen
  \bibfield  {author} {\bibinfo {author} {\bibfnamefont {M.}~\bibnamefont
  {Novello}}, \bibinfo {author} {\bibfnamefont {M.}~\bibnamefont {Visser}},\
  and\ \bibinfo {author} {\bibfnamefont {G.}~\bibnamefont {Volovik}},\ }\href
  {https://books.google.com.br/books?id=-tyXuduShHUC} {\emph {\bibinfo {title}
  {Artificial Black Holes}}}\ (\bibinfo  {publisher} {World Scientific},\
  \bibinfo {year} {2002})\BibitemShut {NoStop}%
\bibitem [{\citenamefont {Pendry}\ \emph {et~al.}(2006)\citenamefont {Pendry},
  \citenamefont {Schurig},\ and\ \citenamefont {Smith}}]{Pendry2006}%
  \BibitemOpen
  \bibfield  {author} {\bibinfo {author} {\bibfnamefont {J.~B.}\ \bibnamefont
  {Pendry}}, \bibinfo {author} {\bibfnamefont {D.}~\bibnamefont {Schurig}},\
  and\ \bibinfo {author} {\bibfnamefont {D.~R.}\ \bibnamefont {Smith}},\
  }\bibfield  {title} {\bibinfo {title} {Controlling electromagnetic fields},\
  }\href {https://doi.org/10.1126/science.1125907} {\bibfield  {journal}
  {\bibinfo  {journal} {Science}\ }\textbf {\bibinfo {volume} {312}},\ \bibinfo
  {pages} {1780} (\bibinfo {year} {2006})}\BibitemShut {NoStop}%
\bibitem [{\citenamefont {Smith}\ \emph {et~al.}(2000)\citenamefont {Smith},
  \citenamefont {Padilla}, \citenamefont {Vier}, \citenamefont {Nemat-Nasser},\
  and\ \citenamefont {Schultz}}]{PhysRevLett.84.4184}%
  \BibitemOpen
  \bibfield  {author} {\bibinfo {author} {\bibfnamefont {D.~R.}\ \bibnamefont
  {Smith}}, \bibinfo {author} {\bibfnamefont {W.~J.}\ \bibnamefont {Padilla}},
  \bibinfo {author} {\bibfnamefont {D.~C.}\ \bibnamefont {Vier}}, \bibinfo
  {author} {\bibfnamefont {S.~C.}\ \bibnamefont {Nemat-Nasser}},\ and\ \bibinfo
  {author} {\bibfnamefont {S.}~\bibnamefont {Schultz}},\ }\bibfield  {title}
  {\bibinfo {title} {Composite medium with simultaneously negative permeability
  and permittivity},\ }\href {https://doi.org/10.1103/PhysRevLett.84.4184}
  {\bibfield  {journal} {\bibinfo  {journal} {Phys. Rev. Lett.}\ }\textbf
  {\bibinfo {volume} {84}},\ \bibinfo {pages} {4184} (\bibinfo {year}
  {2000})}\BibitemShut {NoStop}%
\bibitem [{\citenamefont {Fleischhauer}\ \emph {et~al.}(2005)\citenamefont
  {Fleischhauer}, \citenamefont {Imamoglu},\ and\ \citenamefont
  {Marangos}}]{RevModPhys.77.633}%
  \BibitemOpen
  \bibfield  {author} {\bibinfo {author} {\bibfnamefont {M.}~\bibnamefont
  {Fleischhauer}}, \bibinfo {author} {\bibfnamefont {A.}~\bibnamefont
  {Imamoglu}},\ and\ \bibinfo {author} {\bibfnamefont {J.~P.}\ \bibnamefont
  {Marangos}},\ }\bibfield  {title} {\bibinfo {title} {Electromagnetically
  induced transparency: Optics in coherent media},\ }\href
  {https://doi.org/10.1103/RevModPhys.77.633} {\bibfield  {journal} {\bibinfo
  {journal} {Rev. Mod. Phys.}\ }\textbf {\bibinfo {volume} {77}},\ \bibinfo
  {pages} {633} (\bibinfo {year} {2005})}\BibitemShut {NoStop}%
\bibitem [{\citenamefont {Perlick}(2011)}]{perlick2011}%
  \BibitemOpen
  \bibfield  {author} {\bibinfo {author} {\bibfnamefont {V.}~\bibnamefont
  {Perlick}},\ }\bibfield  {title} {\bibinfo {title} {On the hyperbolicity of
  maxwell's equations with a local constitutive law},\ }\href
  {https://doi.org/10.1063/1.3579133} {\bibfield  {journal} {\bibinfo
  {journal} {Journal of Mathematical Physics}\ }\textbf {\bibinfo {volume}
  {52}},\ \bibinfo {pages} {042903} (\bibinfo {year} {2011})}\BibitemShut
  {NoStop}%
\bibitem [{\citenamefont {Hadamard}(1903)}]{hadamard}%
  \BibitemOpen
  \bibfield  {author} {\bibinfo {author} {\bibfnamefont {J.}~\bibnamefont
  {Hadamard}},\ }\href@noop {} {\emph {\bibinfo {title} {Le\c cons sur la
  propagation des ondes et les \'equations de hydrodynamique}}}\ (\bibinfo
  {publisher} {Hermann},\ \bibinfo {address} {Paris, France},\ \bibinfo {year}
  {1903})\BibitemShut {NoStop}%
\bibitem [{\citenamefont {Zakharov}(1973)}]{zakharov}%
  \BibitemOpen
  \bibfield  {author} {\bibinfo {author} {\bibfnamefont {V.~D.}\ \bibnamefont
  {Zakharov}},\ }\href@noop {} {\emph {\bibinfo {title} {Gravitational waves in
  Einstein's theory}}}\ (\bibinfo  {publisher} {John Wiley \& Sons},\ \bibinfo
  {address} {New York, USA},\ \bibinfo {year} {1973})\BibitemShut {NoStop}%
\bibitem [{\citenamefont {{De Lorenci}}\ and\ \citenamefont
  {Klippert}(2006)}]{DELORENCI200661}%
  \BibitemOpen
  \bibfield  {author} {\bibinfo {author} {\bibfnamefont {V.~A.}\ \bibnamefont
  {{De Lorenci}}}\ and\ \bibinfo {author} {\bibfnamefont {R.}~\bibnamefont
  {Klippert}},\ }\bibfield  {title} {\bibinfo {title} {Electromagnetic light
  rays in local dielectrics},\ }\href
  {https://doi.org/https://doi.org/10.1016/j.physleta.2006.04.010} {\bibfield
  {journal} {\bibinfo  {journal} {Physics Letters A}\ }\textbf {\bibinfo
  {volume} {357}},\ \bibinfo {pages} {61} (\bibinfo {year} {2006})}\BibitemShut
  {NoStop}%
\bibitem [{\citenamefont {{De Lorenci}}\ and\ \citenamefont
  {Goulart}(2008)}]{goulart2008}%
  \BibitemOpen
  \bibfield  {author} {\bibinfo {author} {\bibfnamefont {V.~A.}\ \bibnamefont
  {{De Lorenci}}}\ and\ \bibinfo {author} {\bibfnamefont {G.~P.}\ \bibnamefont
  {Goulart}},\ }\bibfield  {title} {\bibinfo {title} {Magnetoelectric
  birefringence revisited},\ }\href@noop {} {\bibfield  {journal} {\bibinfo
  {journal} {Physics Review D}\ }\textbf {\bibinfo {volume} {78}},\ \bibinfo
  {pages} {045015} (\bibinfo {year} {2008})}\BibitemShut {NoStop}%
\bibitem [{\citenamefont {Bittencourt}\ \emph {et~al.}(2017)\citenamefont
  {Bittencourt}, \citenamefont {Camargo}, \citenamefont {De~Lorenci},\ and\
  \citenamefont {Klippert}}]{PhysRevA.95.033826}%
  \BibitemOpen
  \bibfield  {author} {\bibinfo {author} {\bibfnamefont {E.}~\bibnamefont
  {Bittencourt}}, \bibinfo {author} {\bibfnamefont {G.~H.~S.}\ \bibnamefont
  {Camargo}}, \bibinfo {author} {\bibfnamefont {V.~A.}\ \bibnamefont
  {De~Lorenci}},\ and\ \bibinfo {author} {\bibfnamefont {R.}~\bibnamefont
  {Klippert}},\ }\bibfield  {title} {\bibinfo {title} {Controlled opacity in a
  class of nonlinear dielectric media},\ }\href
  {https://doi.org/10.1103/PhysRevA.95.033826} {\bibfield  {journal} {\bibinfo
  {journal} {Phys. Rev. A}\ }\textbf {\bibinfo {volume} {95}},\ \bibinfo
  {pages} {033826} (\bibinfo {year} {2017})}\BibitemShut {NoStop}%
\bibitem [{\citenamefont {Novello}\ and\ \citenamefont
  {Bittencourt}(2012)}]{edu2012}%
  \BibitemOpen
  \bibfield  {author} {\bibinfo {author} {\bibfnamefont {M.}~\bibnamefont
  {Novello}}\ and\ \bibinfo {author} {\bibfnamefont {E.}~\bibnamefont
  {Bittencourt}},\ }\bibfield  {title} {\bibinfo {title} {Gordon metric
  revisited},\ }\href@noop {} {\bibfield  {journal} {\bibinfo  {journal}
  {Physical Review D}\ }\textbf {\bibinfo {volume} {86}},\ \bibinfo {pages}
  {124024} (\bibinfo {year} {2012})}\BibitemShut {NoStop}%
\bibitem [{\citenamefont {Bittencourt}\ \emph {et~al.}(2016)\citenamefont
  {Bittencourt}, \citenamefont {Pereira}, \citenamefont {Smolyaninov},\ and\
  \citenamefont {Smolyaninova}}]{Bittencourt_2016}%
  \BibitemOpen
  \bibfield  {author} {\bibinfo {author} {\bibfnamefont {E.}~\bibnamefont
  {Bittencourt}}, \bibinfo {author} {\bibfnamefont {J.~P.}\ \bibnamefont
  {Pereira}}, \bibinfo {author} {\bibfnamefont {I.~I.}\ \bibnamefont
  {Smolyaninov}},\ and\ \bibinfo {author} {\bibfnamefont {V.~N.}\ \bibnamefont
  {Smolyaninova}},\ }\bibfield  {title} {\bibinfo {title} {The flexibility of
  optical metrics},\ }\href {https://doi.org/10.1088/0264-9381/33/16/165008}
  {\bibfield  {journal} {\bibinfo  {journal} {Classical and Quantum Gravity}\
  }\textbf {\bibinfo {volume} {33}},\ \bibinfo {pages} {165008} (\bibinfo
  {year} {2016})}\BibitemShut {NoStop}%
\bibitem [{\citenamefont {{De Lorenci}}\ and\ \citenamefont
  {Klippert}(2002)}]{delorenci2002}%
  \BibitemOpen
  \bibfield  {author} {\bibinfo {author} {\bibfnamefont {V.~A.}\ \bibnamefont
  {{De Lorenci}}}\ and\ \bibinfo {author} {\bibfnamefont {R.}~\bibnamefont
  {Klippert}},\ }\bibfield  {title} {\bibinfo {title} {Analogue gravity from
  electrodynamics in nonlinear media},\ }\href@noop {} {\bibfield  {journal}
  {\bibinfo  {journal} {Physical Review D}\ }\textbf {\bibinfo {volume} {65}},\
  \bibinfo {pages} {064027} (\bibinfo {year} {2002})}\BibitemShut {NoStop}%
\bibitem [{\citenamefont {{De Lorenci}}(2002)}]{delorenci2002b}%
  \BibitemOpen
  \bibfield  {author} {\bibinfo {author} {\bibfnamefont {V.~A.}\ \bibnamefont
  {{De Lorenci}}},\ }\bibfield  {title} {\bibinfo {title} {Effective geometry
  for light traveling in material media},\ }\href@noop {} {\bibfield  {journal}
  {\bibinfo  {journal} {Physical Review E}\ }\textbf {\bibinfo {volume} {65}},\
  \bibinfo {pages} {026612} (\bibinfo {year} {2002})}\BibitemShut {NoStop}%
\bibitem [{\citenamefont {Morris}\ and\ \citenamefont {Thorne}(1988)}]{morris}%
  \BibitemOpen
  \bibfield  {author} {\bibinfo {author} {\bibfnamefont {M.~S.}\ \bibnamefont
  {Morris}}\ and\ \bibinfo {author} {\bibfnamefont {K.~S.}\ \bibnamefont
  {Thorne}},\ }\bibfield  {title} {\bibinfo {title} {Wormholes in spacetime and
  their use for interstellar travel: a tool for teaching general relativity},\
  }\href@noop {} {\bibfield  {journal} {\bibinfo  {journal} {American Journal
  of Physics}\ }\textbf {\bibinfo {volume} {56}},\ \bibinfo {pages} {395}
  (\bibinfo {year} {1988})}\BibitemShut {NoStop}%
\bibitem [{\citenamefont {R\"atzel}\ \emph {et~al.}(2011)\citenamefont
  {R\"atzel}, \citenamefont {Rivera},\ and\ \citenamefont
  {Schuller}}]{PhysRevD.83.044047}%
  \BibitemOpen
  \bibfield  {author} {\bibinfo {author} {\bibfnamefont {D.}~\bibnamefont
  {R\"atzel}}, \bibinfo {author} {\bibfnamefont {S.}~\bibnamefont {Rivera}},\
  and\ \bibinfo {author} {\bibfnamefont {F.~P.}\ \bibnamefont {Schuller}},\
  }\bibfield  {title} {\bibinfo {title} {Geometry of physical dispersion
  relations},\ }\href {https://doi.org/10.1103/PhysRevD.83.044047} {\bibfield
  {journal} {\bibinfo  {journal} {Phys. Rev. D}\ }\textbf {\bibinfo {volume}
  {83}},\ \bibinfo {pages} {044047} (\bibinfo {year} {2011})}\BibitemShut
  {NoStop}%
\bibitem [{\citenamefont {GÅRDING}(1959)}]{10.2307/24900665}%
  \BibitemOpen
  \bibfield  {author} {\bibinfo {author} {\bibfnamefont {L.}~\bibnamefont
  {GÅRDING}},\ }\bibfield  {title} {\bibinfo {title} {An inequality for
  hyperbolic polynomials},\ }\href {http://www.jstor.org/stable/24900665}
  {\bibfield  {journal} {\bibinfo  {journal} {Journal of Mathematics and
  Mechanics}\ }\textbf {\bibinfo {volume} {8}},\ \bibinfo {pages} {957}
  (\bibinfo {year} {1959})}\BibitemShut {NoStop}%
\bibitem [{\citenamefont {Beig}(2006)}]{Beig2006}%
  \BibitemOpen
  \bibfield  {author} {\bibinfo {author} {\bibfnamefont {R.}~\bibnamefont
  {Beig}},\ }\bibinfo {title} {Concepts of hyperbolicity and relativistic
  continuum mechanics},\ in\ \href {https://doi.org/10.1007/3-540-33484-X_5}
  {\emph {\bibinfo {booktitle} {Analytical and Numerical Approaches to
  Mathematical Relativity}}},\ \bibinfo {editor} {edited by\ \bibinfo {editor}
  {\bibfnamefont {J.}~\bibnamefont {Frauendiener}}, \bibinfo {editor}
  {\bibfnamefont {D.~J.}\ \bibnamefont {Giulini}},\ and\ \bibinfo {editor}
  {\bibfnamefont {V.}~\bibnamefont {Perlick}}}\ (\bibinfo  {publisher}
  {Springer Berlin Heidelberg},\ \bibinfo {address} {Berlin, Heidelberg},\
  \bibinfo {year} {2006})\ pp.\ \bibinfo {pages} {101--116}\BibitemShut
  {NoStop}%
\end{thebibliography}%
\end{document}